\documentclass[11pt,a4paper]{article}
\usepackage[T1]{fontenc}
\usepackage[utf8]{inputenc}
\usepackage[a4paper,margin=1in,footskip=10mm]{geometry}
\usepackage{amsmath,amssymb,mathtools}
\usepackage{newtxtext,newtxmath}

\usepackage{graphicx,booktabs,array}
\usepackage{microtype}
\usepackage{enumitem}
\usepackage{caption}
\usepackage{placeins}
\usepackage{needspace}
\usepackage{titlesec}
\usepackage{xurl}
\usepackage{xcolor}
\usepackage{hyperref}
\hypersetup{hidelinks,bookmarksnumbered=true,
 pdftitle={Biologically Inspired Predictive Gating Signed-Dyadic Coding for Mandarin Keyword Spotting},
 pdfauthor={Zijian Zhou, Rui Liu, Rongxiang Nie, Rui Ma, Ziyao Zhang, Xin Du},pdfsubject={Predictive Gating Signed-Dyadic Coding for Mandarin keyword spotting}}
\titleformat{\section}{\large\bfseries}{\thesection}{.7em}{}
\titleformat{\subsection}{\normalsize\bfseries}{\thesubsection}{.7em}{}
\titleformat{\subsubsection}{\normalsize\itshape}{\thesubsubsection}{.7em}{}
\titlespacing*{\section}{0pt}{13pt plus 2pt minus 1pt}{5pt}
\titlespacing*{\subsection}{0pt}{10pt plus 2pt minus 1pt}{3pt}
\titlespacing*{\subsubsection}{0pt}{8pt plus 1pt minus 1pt}{2pt}
\setlist{leftmargin=1.6em,itemsep=3pt,topsep=4pt,parsep=0pt}

\title{\bfseries Biologically Inspired Predictive Gating\\[3pt]
Signed-Dyadic Coding for Mandarin Keyword Spotting}
\author{%
  \normalsize Zijian Zhou\textsuperscript{1},
  Rui Liu\textsuperscript{1,*},
  Rongxiang Nie\textsuperscript{1},
  Rui Ma\textsuperscript{1},
  Ziyao Zhang\textsuperscript{1},
  Xin Du\textsuperscript{2,*}\\[7pt]
  \small \textsuperscript{1}\,Sivy Limited, Hangzhou, China\\[3pt]
  \small \textsuperscript{2}\,The Cavendish Laboratory, Department of Physics,\\
  \small University of Cambridge, Cambridge, UK\\[6pt]
  \small \textsuperscript{*}\,Corresponding authors.
}
\date{}

\begin{document}
\maketitle

\begin{abstract}
In continuous spiking keyword spotting, temporally correlated acoustic frames can repeatedly trigger memory access and neural computation at a dense input interface, even when hidden activity is sparse. Predictive processing and sensory adaptation suggest retaining predictable input locally and responding selectively to deviations. Inspired by this mechanism, we propose \textbf{Predictive Gating Signed-Dyadic Coding (PG)}. PG maintains a local prediction for each input channel and transmits an event only when the residual exceeds a learned threshold. Each event carries a sign and a finite power-of-two magnitude, and the receiver reconstructs it with fixed-point shifts and additions. In Mandarin keyword spotting, INPUT-PG reaches \textbf{90.00\%} accuracy while reducing mean SRAM reads by \textbf{33.87\%} and inference cycles by \textbf{32.77\%}. PG provides a hardware interface for research on low-power continuous speech processing and event-based sensory systems.
\end{abstract}

\FloatBarrier
\section{Introduction}\label{introduction}

Continuous keyword spotting processes an acoustic stream one frame at a time under tight energy and latency budgets. Adjacent log-mel frames are temporally correlated, but a dense input interface still transfers every frame and repeatedly triggers memory access and downstream computation. The input interface therefore adds work even when hidden activity is sparse. Biological predictive processing, efficient coding, and sensory adaptation describe related mechanisms in which a system maintains an internal estimate, attenuates predictable input, and responds selectively to deviations \cite{ref1,ref2,ref7,ref8,ref16,ref21}. We translate this mechanism into a local-state encoder that preserves predictable content and transmits significant mismatches.

Existing event encoders expose different trade-offs. Rate coding represents amplitude through repeated events over a time window, and TTFS coding represents it through spike latency \cite{ref13,ref17}. Delta-style encoding uses changes from a reference state, but applying the same operation to already thresholded hidden spikes can create events for both rising and falling transitions. The placement of an event code therefore matters as much as the code itself.

We propose \textbf{Predictive Gating Signed-Dyadic Coding (PG)}, a trainable lossy event code for the acoustic input. PG maintains a local prediction for each channel, computes the residual from the current frame, and emits an event only when the residual crosses a learned threshold. Each event carries a sign and a magnitude selected from a finite dyadic alphabet. The receiving datapath reconstructs each update with fixed-point comparisons, shifts, and additions, yielding a multiplier-free interface.

We evaluate Mandarin keyword spotting with three deployment schemes. DIRECT encodes every input frame, INPUT-PG applies PG only at the acoustic input, and ALL-PG applies PG at the input and hidden boundaries. On 720 test clips, INPUT-PG reaches 90.00\% accuracy while reducing mean SRAM reads by 33.87\% and inference cycles by 32.77\%. A routed evaluation on one fixed clip reports a 39.49\% reduction in incremental inference energy, excluding weight loading. Software workload is evaluated over 720 test clips, whereas routed energy is evaluated on one fixed clip.

The resulting study identifies selective input prediction as the useful deployment point for this task. It connects temporal redundancy in acoustic frames with measurable reductions in memory traffic and execution work, and provides a hardware interface for low-power continuous speech processing and event-based sensory systems.

\begin{figure}[!htbp]
\centering
\makebox[\linewidth][c]{\includegraphics[width=174mm,height=.66\textheight,keepaspectratio]{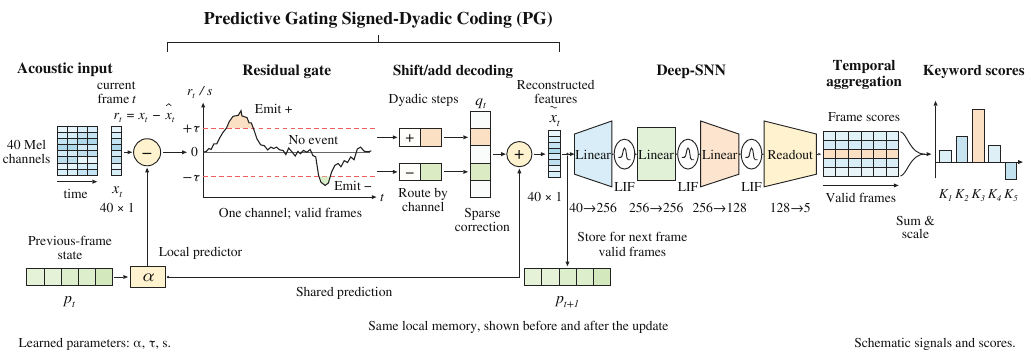}}
\caption{Predictive Gating Signed-Dyadic Coding (PG) and its interface with keyword classification. PG maintains a shared prediction, transmits thresholded signed-dyadic residuals, and reconstructs the input before the Deep-SNN. Temporal aggregation produces the keyword scores.}
\label{fig:1}
\end{figure}

Our contributions are the following.

\begin{enumerate}
\def\labelenumi{\arabic{enumi}.}
\item
  \textbf{Predictive Gating Signed-Dyadic Coding (PG)}, a trainable lossy event code that maintains a local prediction, suppresses predictable input, and represents threshold-crossing residuals with a finite signed-dyadic alphabet for fixed-point reconstruction;
\item
  \textbf{A selective placement principle for predictive event coding}, evaluated through DIRECT, INPUT-PG, and ALL-PG across a feedforward Deep-SNN and a causal LIF-TCN. The comparison shows why prediction is useful where temporal redundancy remains, whereas coding an already thresholded hidden stream can add workload.
\item
  \textbf{A 720-clip integer evaluation} that jointly measures task accuracy, input activity, SRAM reads, and inference cycles. INPUT-PG reaches 90.00\% accuracy and reduces mean SRAM reads and inference cycles by 33.87\% and 32.77\%.
\item
  \textbf{A routed Sky130HD evaluation} that measures a 39.49\% reduction in incremental inference energy for one fixed hardware clip, excluding weight loading. Software accuracy and workload are evaluated separately over 720 test clips.
\end{enumerate}

\FloatBarrier
\section{Related Work}\label{related-work}

\subsection{Biological foundations}\label{biological-foundations}

Biological sensory systems adapt their responses to the temporal statistics of stimulation. Predictive processing proposes that neural circuits maintain an internal estimate of incoming signals and emphasize deviations from that estimate \cite{ref8,ref16}. Sensory adaptation produces a weaker response to persistent stimulation and restores responsiveness when the stimulus changes \cite{ref7,ref21}. Efficient coding provides a complementary account in which limited neural resources are preferentially allocated to stimulus components that are less predictable from their statistics \cite{ref1,ref2,ref11}.

Together, these mechanisms describe a biological response pattern in which an internal estimate is maintained, predictable stimulation is attenuated, and deviations receive stronger responses. This pattern provides the biological principle for PG.

\subsection{Hardware-efficient event coding}\label{hardware-efficient-event-coding}

SNN accelerators exploit sparse hidden spikes, but a dense acoustic interface can still repeatedly move frames and trigger input-side work. The hardware benefit of an event code therefore depends on its payload, state, routing, and decoder cost, in addition to event count. We report communication activity, digital workload, and routed physical energy as separate measures.

Existing encoders expose different trade-offs. DeltaKWS suppresses changes in acoustic features and recurrent states to avoid redundant updates \cite{ref10}. Rate coding represents amplitude through repeated events over a time window, whereas TTFS represents it by spike latency. Our RATE-T8 and TTFS-T8 baselines replay quantized amplitudes at the input while retaining the frame loop. Sigma--Delta modulation shapes quantization error through an integrator and feedback loop \cite{ref19}. PG keeps an explicit local prediction and sends an addressable signed update from a finite dyadic alphabet once the residual crosses a threshold.

PG is applied at the acoustic input, where adjacent log mel frames retain temporal redundancy before entering the SNN. Hidden LIF boundaries already expose thresholded binary events, so PG incurs packet, state, and decoder costs on an already sparse stream. We compare input and hidden placement and count event, SRAM, cycle, and decoder costs for each.
\FloatBarrier
\section{Method}\label{method}

We define the method through four linked components. They are a biological computational abstraction, a selective event interface, a hardware datapath, and a measurement protocol. Each component is evaluated under a common deployment contract for its event, memory, and energy consequences.

\subsection{Biologically inspired abstraction of prediction errors}\label{bio-inspired-predictive-error-abstraction}

We abstract these mechanisms as a local-state predictor that retains predictable sensory content and emits a sparse directed update for an unexpected change. The following equations define this abstraction in accelerator-level quantities.

For each encoded boundary, time step \(t\), and address \(a\), let \(x_{t,a}\) be the signed input and \(p_{t,a}\) its synchronized local prediction state, initialized by \(p_{0,a} = 0\). The trainable parameters are constrained as \(s=\operatorname{softplus}(\theta_s)+10^{-6}\), \(\tau=\operatorname{sigmoid}(\theta_\tau)\), and \(\alpha=\operatorname{sigmoid}(\theta_\alpha)\). A leaky prediction and residual are

\begin{equation}
\label{eq:1}
\widehat{x}_{t,a}=\alpha p_{t,a},
\qquad r_{t,a}=x_{t,a}-\widehat{x}_{t,a}.
\end{equation}

Here \(s > 0\) is the learned scale, \(\tau \in (0,1)\) is the normalized threshold, and \(\alpha \in (0,1)\) is the predictor coefficient; each is frozen as a fixed point constant in the RTL experiment. With \(\mathcal{K}=\{0,1,\ldots,K_{\max}\}\) denoting the available exponent slots, the normalized magnitude and hard gate are

\begin{equation}
\label{eq:2}
u_{t,a}=\frac{|r_{t,a}|}{s},
\qquad g_{t,a}=\mathbf{1}\!\left[u_{t,a}\geq\tau\right].
\end{equation}

For every active address, exponent selection and decoded residual are

\begin{equation}
\label{eq:3}
\begin{aligned}
k_{t,a} &= \operatorname{clip}_{\mathcal{K}}\!\left(
\operatorname{round}\!\left[-\log_2\!\left(\max(u_{t,a},10^{-12})\right)\right]\right),\\
q_{t,a} &= g_{t,a}\operatorname{sgn}(r_{t,a})\,s\,2^{-k_{t,a}}.
\end{aligned}
\end{equation}

When \(g_{t,a} = 1\), the transmitted packet is \(e_{t,a}=(a,\operatorname{sgn}(r_{t,a}),k_{t,a})\); otherwise no packet is emitted for that address. Let \(m_{t} \in \{ 0,1\}\) indicate a valid frame that is not padding. The reconstructed output and state update are

\begin{equation}
\label{eq:4}
\left(\widetilde{x}_{t,a},p_{t+1,a}\right)=
\begin{cases}
\left(\widehat{x}_{t,a}+q_{t,a},\widehat{x}_{t,a}+q_{t,a}\right), & m_t=1,\\
\left(0,p_{t,a}\right), & m_t=0.
\end{cases}
\end{equation}

For a valid address whose normalized residual is below the threshold, \(q_{t,a}=0\) and the state advances through the shared predictor. A padding frame emits zero and holds the previous state. The forward path suppresses subthreshold residuals and advances the state through the shared predictor. During training, the gate uses \(g_{t,a}^{\mathrm{soft}}=\operatorname{sigmoid}(12(u_{t,a}-\tau))\) and the straight-through value \(g_{t,a}^{\mathrm{ST}}=g_{t,a}^{\mathrm{soft}}+\operatorname{stopgrad}(g_{t,a}-g_{t,a}^{\mathrm{soft}})\). Exponent selection uses the analogous straight-through construction with \(r_{t,a}/s\) as its backward surrogate \cite{ref14}. The forward computation uses the hard event representation defined above.

\subsection{Signed-dyadic payload and hardware compatibility}\label{signed-dyadic-payload-and-hardware-compatibility}

The sign preserves the direction of a change, while the finite exponent set preserves a coarse magnitude hierarchy. Since \(2^{- k}\) is a shift, decoding can be implemented by sign extension, shift selection, and add/subtract. The resulting datapath uses fixed-point operations instead of a general multiplier, floating-point exponentiation, or a large lookup table. The number of exponent slots, scale, threshold, and predictor leak are explicit knobs for the accuracy--activity--energy frontier. These parameters implement selective responses with multiple magnitudes in the hardware datapath.

For INPUT-PG trained from scratch, the learned scale, normalized threshold, and predictor coefficient average \((0.2118,0.2306,0.8917)\) for Deep-SNN and \((0.2186,0.2321,0.8949)\) for LIF-TCN across the three seeds. The routed Deep-SNN assets freeze these quantities as \(\text{scale\_q15} = 6863\), \(\text{threshold\_q15} = 7472\), and \(\text{alpha\_q16} = 58629\). The same constants are used during integer inference.

\subsection{Selective placement}\label{selective-placement}

We compare three placement schemes. DIRECT encodes every input frame densely and uses ordinary LIF spikes at the hidden boundaries. INPUT-PG applies PG only at the acoustic input. ALL-PG applies PG at both the input and hidden boundaries.

INPUT-PG is the primary design because the input is both temporally correlated and not yet explicitly sparse. Hidden LIF spikes have already undergone a thresholding operation, so encoding their transitions can add traffic. A layer \(\ell\) is therefore a candidate for PG only if the expected savings exceed the packet, state, and decoder cost. We express this condition as

\begin{equation}
\label{eq:5}
\mathbb{E}\!\left[C_{\mathrm{encoded},\ell}\right]+C_{\mathrm{PG},\ell}
<\mathbb{E}\!\left[C_{\mathrm{ordinary},\ell}\right].
\end{equation}

\begin{figure}[!htbp]
\centering
\makebox[\linewidth][c]{\includegraphics[width=174mm,height=.66\textheight,keepaspectratio]{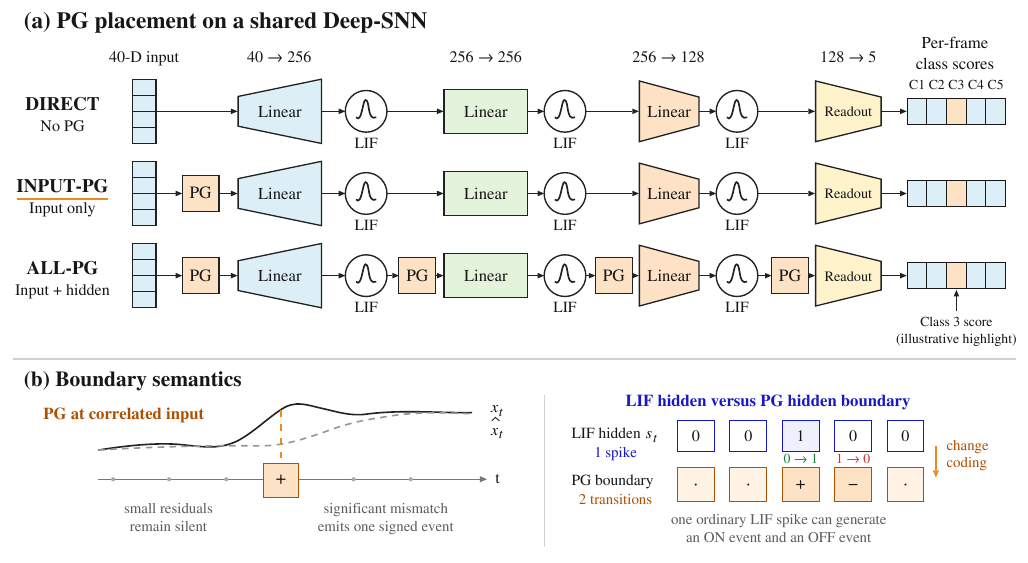}}
\caption{Selective PG placement and boundary semantics. DIRECT has no PG, INPUT-PG places PG before the first linear map, and ALL-PG places PG before all learned maps. Small residuals remain silent, whereas significant mismatches emit signed events; hidden LIF transitions are shown schematically.}
\label{fig:2}
\end{figure}

Equation~\eqref{eq:5} gives the placement criterion used to interpret the three variants. Figure~\ref{fig:2} compares the placements and illustrates their different boundary semantics.

\subsection{Accelerator architecture and hardware verification}\label{accelerator-architecture-and-hardware-verification}

\subsubsection{Event packet and decoder datapath}\label{event-packet-and-decoder-datapath}

Each PG event contains an address, a valid bit, a sign, and an exponent. The decoder sign extends the payload, shifts it by the selected exponent, and accumulates it into a predictor register. These operations implement the prediction update with integer arithmetic. The routed accounting includes event routing, state registers, comparator activity, clocking, SRAM reads, and decoder activity.

\subsubsection{Routed implementation and energy accounting}\label{routed-implementation-and-energy-accounting}

We map the four-layer Deep-SNN with the Sky130HD library at the typical corner, 25 \ensuremath{^{\circ}}C, 1.8 V, and a nominal 10 MHz clock. From the routed ODB and SPEF files, we generate functional gate-level FST activity without propagation delays, convert all nets to hierarchical SAIF, and annotate OpenROAD power. The reported runs annotate 100\% of nets; internal driver-pin coverage is at least 90.87\%, and SPEF coverage is at least 96.87\%.

Resident energy assumes that weights are already in SRAM. Transported energy adds the fixed 220,948-byte weight load. The energy boundary excludes the analog front end, ADC, log mel extraction, package, off-chip memory, and system I/O. Let \(n_{\mathrm{start},v}\) and \(n_{\mathrm{stop},v}\) delimit the inference phase of variant \(v\) after parameter loading. Let \(r_{v,p}\lbrack n\rbrack\) denote the read enable of logical SRAM port \(p\). The measured work quantities are

\begin{equation}
\label{eq:6}
\begin{aligned}
C_v &= n_{\mathrm{stop},v}-n_{\mathrm{start},v},
\qquad T_{\mathrm{infer},v}=\frac{C_v}{f_{\mathrm{clk}}},\\
N_{\mathrm{read},v} &= \sum_{n=n_{\mathrm{start},v}}^{n_{\mathrm{stop},v}-1}
\sum_{p\in\mathcal{P}}\mathbf{1}\!\left[r_{v,p}[n]=1\right].
\end{aligned}
\end{equation}

In the RTL, the read counter increments once in each state that reads a weight, a hidden bias, or an output bias. With average dynamic powers \(P_{\mathrm{infer},v}\), \(P_{\mathrm{load},v}\), and \(P_{\mathrm{idle},v}\) measured over their respective windows,

\begin{equation}
\label{eq:7}
\begin{aligned}
E_{\mathrm{active},v} &= P_{\mathrm{infer},v}T_{\mathrm{infer},v},\\
E_{\mathrm{resident},v} &= \left(P_{\mathrm{infer},v}-P_{\mathrm{idle},v}\right)T_{\mathrm{infer},v},\\
E_{\mathrm{load,inc},v} &= \left(P_{\mathrm{load},v}-P_{\mathrm{idle},v}\right)T_{\mathrm{load},v},\\
E_{\mathrm{transported},v} &= E_{\mathrm{load,inc},v}+E_{\mathrm{resident},v}.
\end{aligned}
\end{equation}

Subtracting the idle baseline isolates incremental dynamic work, while active energy retains the physical activity window. The routed table uses seed 17 and clip 3826 with 232 valid frames. All five variants use the same SRAM organization, weight image size, variant-specific weights, and PVT conditions.

\subsubsection{RTL consistency and implementation status}\label{rtl-consistency-and-implementation-status}

The integer software evaluator matches RTL simulations on 8-frame and 64-frame prefixes and on the complete 232-frame test clip. The checks compare logits, predictions, event counts, SRAM reads, and execution cycles. Software evaluation covers all 720 test clips. Routed measurements use one frozen clip, while integer accuracy and workload means use the full test set. The routed run is reported as a prototype evaluation with the measured timing and DRC status.

\subsection{Experimental protocol and measurement methodology}\label{experimental-protocol-and-measurement-methodology}

\subsubsection{Task and data}\label{task-and-data}

The corpus contains 4,545 Mandarin utterance clips from 101 speakers and five keyword classes. The training, validation, and test partitions contain 3,150, 675, and 720 clips, respectively, with no speaker overlap between partitions. We resample audio from 48 kHz to 16 kHz and represent it with 40-channel log mel features over at most 560 frames. Normalization uses training-set statistics, and padding frames are excluded from event and activity counts. Each test clip contains an average of 248.311 valid frames, or 9,932.444 dense feature values.

\subsubsection{Networks and fair training}\label{networks-and-fair-training}

The Deep-SNN has \(40 \rightarrow 256 \rightarrow 256 \rightarrow 128 \rightarrow 5\) linear layers with three streaming LIF hidden layers, \(\beta = 0.9\), threshold 1.0, and subtractive reset (109,841 parameters). The LIF-TCN has four causal residual Conv1d/LIF blocks with dilations 1, 2, 4, and 8, masked temporal pooling, and a five-output classifier (99,284 parameters). DIRECT, INPUT-PG, and ALL-PG are each trained from scratch for 50 epochs with seeds 7, 17, and 27. Training uses batch size 128, learning rate \(10^{- 3}\), weight decay \(10^{- 4}\), six exponent slots, and cross-entropy loss. Validation accuracy selects checkpoints, and the test set is not used for epoch selection. Training runs on the device identified as \texttt{cuda:0}. For each architecture and seed, the corresponding variants share the same initialization fingerprint before PG is applied.

We additionally train RATE-T8 and TTFS-T8 on the same Deep-SNN to compare input coding at the scale of the main network. The data partitions are unchanged and contain no speaker overlap. Training uses 50 epochs, batch size 128, AdamW with learning rate 0.001, weight decay 0.0001, and seeds 7, 17, and 27. Checkpoints are selected by validation accuracy. Each new run shares the corresponding DIRECT initialization fingerprint before PG is applied. DIRECT checkpoints serve as the controls. Both encoders operate only at the input; the architecture, acoustic-frame loop, and hidden LIF dynamics remain unchanged.

RATE-T8 replaces a signed feature with a quantized amplitude at magnitude levels 0--8 and counts that level as the number of input events. TTFS-T8 uses magnitude levels 0--7 and counts one event for each nonzero level. In software, both replay signed quantized amplitudes. In RTL, RATE-T8 replays signed unit contributions through the input synapses, whereas TTFS-T8 sends one signed decoded amplitude per active channel. Both encoders retain the original frame loop and operate at the input boundary.

\subsubsection{Integer inference protocol}\label{integer-inference-protocol}

The integer evaluator uses the frozen seed-17 Deep-SNN weights on all 720 test clips from the same split. The partitions have no speaker overlap. Features and PG state use 15 fractional bits, weights use 9 fractional bits, and membrane accumulation uses 24 fractional bits with a leak coefficient represented with 16 fractional bits. PG uses six dyadic exponent slots and compares full integer products in the threshold test. Prediction, residual, and accumulator operations retain their specified signed widths. Input activity counts dense feature values for DIRECT and emitted input events for PG. The evaluator counts SRAM reads for weights and biases and computes inference cycles from the event schedule, excluding parameter loading. Workloads are reported as arithmetic means per clip.

\subsubsection{Activity and statistical conventions}\label{activity-and-statistical-conventions}

For variant \(v\), seed \(r\), and clip \(c\), let \(T_{c}\) be the number of valid frames, \(D = 40\) the input dimension, \(\mathcal{B}_{v}^{\mathrm{PG}}\) the set of boundaries encoded by PG, and \(\mathcal{B}_{v}^{\mathrm{LIF}}\) the set of unencoded hidden LIF boundaries. With \(g_{v,r,c,t,a}^{(b)}\) the PG valid bit and \(s_{v,r,c,t,j}^{(\ell)}\in\{0,1\}\) the hidden spike, the complete communication proxy for one clip is

\begin{equation}
\label{eq:8}
\begin{aligned}
A_{v,r,c} ={}& DT_c\,\mathbf{1}\!\left[v=\mathrm{DIRECT}\right]\\
&+\sum_{b\in\mathcal{B}_v^{\mathrm{PG}}}\sum_{(t,a)\in\mathcal{V}_{c,b}}g_{v,r,c,t,a}^{(b)}\\
&+\sum_{\ell\in\mathcal{B}_v^{\mathrm{LIF}}}\sum_{(t,j)\in\mathcal{V}_{c,\ell}}s_{v,r,c,t,j}^{(\ell)},
\end{aligned}
\end{equation}

where \(\mathcal{V}_{c,b}\) contains only valid time--address pairs, excluding padding. Under this definition, \textsc{DIRECT} counts dense input values and ordinary hidden spikes, \textsc{INPUT-PG} counts input PG events and ordinary hidden spikes, and \textsc{ALL-PG} counts PG events at every encoded boundary. This quantity is an activity proxy. For \(n\) seeds and \(N\) test clips, its reported mean and relative change are

\begin{equation}
\label{eq:9}
\overline{A}_v=\frac{1}{nN}\sum_{r=1}^{n}\sum_{c=1}^{N}A_{v,r,c},
\qquad
\Delta A_v=100\,\frac{\overline{A}_v-\overline{A}_{\mathrm{DIRECT}}}{\overline{A}_{\mathrm{DIRECT}}}.
\end{equation}

Input-boundary activity uses the same equations after restricting Eq.~\eqref{eq:8} to the input term. A negative \(\Delta A_{v}\) denotes a reduction, whose positive magnitude is \(R_{A,v}=-\Delta A_{v}\).

For RATE-T8 and TTFS-T8, the complete proxy is the number of encoded input events plus positive spikes from all three hidden LIF layers. This replaces the dense-input term in Eq.~\eqref{eq:8} and adds no hidden PG boundary. All counts exclude padding and use the same clip and seed averaging in Eq.~\eqref{eq:9}.

Test accuracy for seed \(r\) is the percentage of correctly classified test clips. Its mean and sample standard deviation across seeds are

\begin{equation}
\label{eq:10}
\begin{aligned}
a_{v,r} &= \frac{100}{N}\sum_{c=1}^{N}\mathbf{1}\!\left[\widehat{y}_{v,r,c}=y_c\right],
\qquad \overline{a}_v=\frac{1}{n}\sum_{r=1}^{n}a_{v,r},\\
s_{a,v} &= \sqrt{\frac{1}{n-1}\sum_{r=1}^{n}\left(a_{v,r}-\overline{a}_v\right)^2}.
\end{aligned}
\end{equation}

\begin{sloppypar}
For a paired comparison with \textsc{DIRECT}, \(d_r=a_{v,r}-a_{\mathrm{DIRECT},r}\) in percentage points. With \(\overline{d}=n^{-1}\sum_r d_r\) and \(s_d^2=(n-1)^{-1}\sum_r(d_r-\overline{d})^2\), the reported two-sided 95\% interval is
\end{sloppypar}

\begin{equation}
\label{eq:11}
\overline{d}\ \pm\ t_{0.975,n-1}\frac{s_d}{\sqrt{n}}.
\end{equation}

With \(n=3\), these intervals describe uncertainty across seeds. For any workload or routed metric \(m\in\{C,N_{\mathrm{read}},P,E,\mathrm{area}\}\), the tables use the same baseline convention,

\begin{equation}
\label{eq:12}
\Delta m_v=100\,\frac{m_v-m_{\mathrm{DIRECT}}}{m_{\mathrm{DIRECT}}},
\qquad R_{m,v}=-\Delta m_v,
\end{equation}

Routed measurements use one frozen clip; integer workload means use all 720 test clips.

\subsubsection{Evaluation tracks}\label{evaluation-tracks}

Table~\ref{tab:1} summarizes the evaluation tracks. From-scratch training measures accuracy and communication activity across two architectures and three seeds. Integer inference measures accuracy and mean workload with frozen Deep-SNN weights on all 720 test clips. The routed experiment measures energy for five implementations on one fixed clip. The movable-SRAM study evaluates an additional accuracy--energy frontier on a separate microarchitecture.

\begin{table}[!htbp]
\centering
\caption{Evaluation tracks and auxiliary study. Each row lists its population, primary outputs, and scope.}
\label{tab:1}
\begingroup
\fontsize{9}{11.6}\selectfont
\setlength{\tabcolsep}{3.3pt}
\renewcommand{\arraystretch}{1.2}
\begin{tabular}{@{}>{\raggedright\arraybackslash}p{0.14000\dimexpr\linewidth-8\tabcolsep\relax}>{\raggedright\arraybackslash}p{0.16000\dimexpr\linewidth-8\tabcolsep\relax}>{\raggedright\arraybackslash}p{0.18500\dimexpr\linewidth-8\tabcolsep\relax}>{\raggedright\arraybackslash}p{0.18000\dimexpr\linewidth-8\tabcolsep\relax}>{\raggedright\arraybackslash}p{0.33500\dimexpr\linewidth-8\tabcolsep\relax}@{}}
\toprule
\textbf{Track}
 & \textbf{Question}
 & \textbf{Population}
 & \textbf{Primary outputs}
 & \textbf{Scope}
 \\
\midrule
E1 from scratch, 50 epochs & Task performance & 2 architectures, 3 seeds, 720 clips & accuracy, activity proxy & software; rate/TTFS on Deep-SNN \\
E5 routed & Energy cost of work & Deep-SNN, seed 17, clip 3826 & cycles, reads, power, energy & one fixed clip, after routing \\
Integer inference & Accuracy and workload & Deep-SNN, seed 17, 720 clips & accuracy, cycles, reads & integer software; frozen weights \\
Small movable & Separate frontier & \(40 \rightarrow 64 \rightarrow 5\), seed 17 & Q9 accuracy, routed energy & separate microarchitecture \\
\bottomrule
\end{tabular}
\endgroup
\end{table}

\FloatBarrier
\section{Experimental results}\label{experimental-results}

\subsection{Accuracy and activity across architectures}\label{accuracy-and-activity-under-matched-training}

\begin{figure}[!htb]
\centering
\includegraphics[width=147.9mm,height=.561\textheight,keepaspectratio]{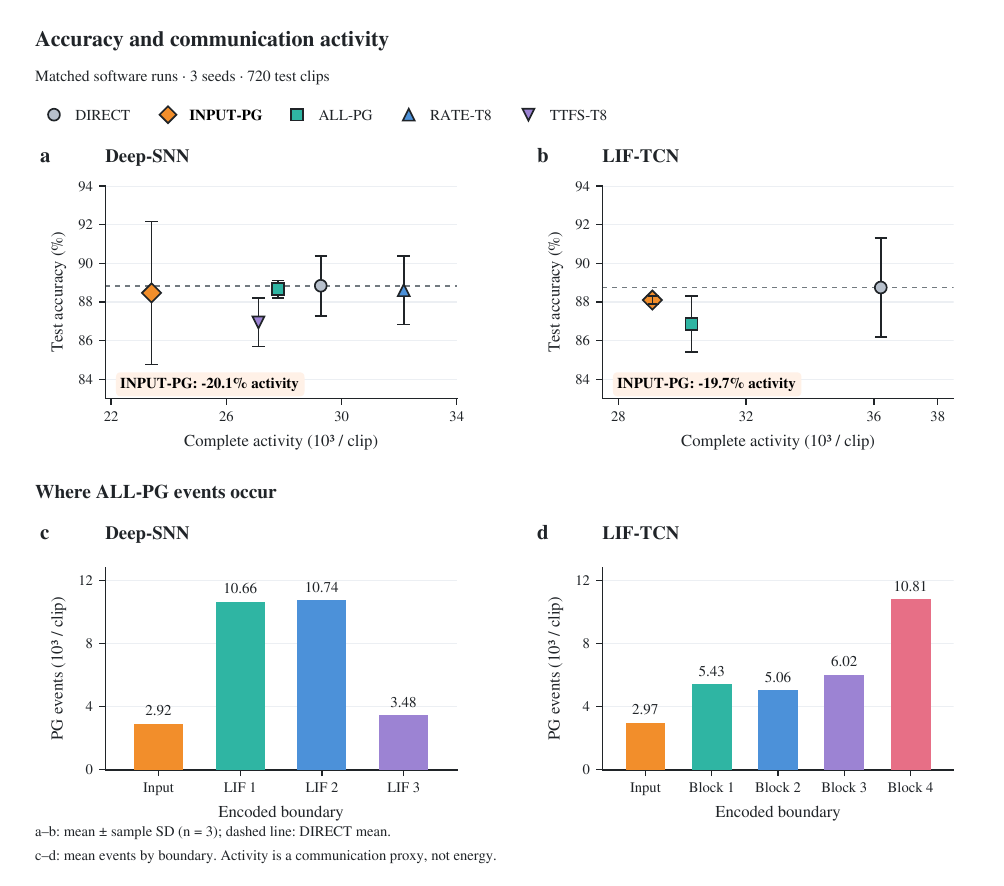}
\caption{Software accuracy and complete communication activity. Results use 50-epoch training over seeds 7, 17 and 27 on 720 test clips; whiskers show sample SD and dashed lines show DIRECT accuracy. RATE-T8 and TTFS-T8 are input-only Deep-SNN baselines. Panels (c-d) show mean ALL-PG events by boundary. Complete activity follows Eqs.~\eqref{eq:8}--\eqref{eq:9} and is reported as a communication proxy.}
\label{fig:3}
\end{figure}

Across both architectures, INPUT-PG keeps the observed accuracy close to DIRECT while reducing communication activity (Table~\ref{tab:2}). Its mean accuracy is 88.47\% for Deep-SNN and 88.10\% for LIF-TCN, compared with 88.84\% and 88.75\% for DIRECT. Input events fall from 9,932.444 dense values per clip to 2,806.5 and 2,684.8, giving reductions of 71.74\% and 72.97\%, respectively. Complete communication activity falls by 20.07\% and 19.74\%. Across the three seeds, the paired intervals are compatible with the DIRECT values, while input and complete activity are consistently lower. Figure~\ref{fig:3}(a-b) shows the accuracy and activity comparison.

On Deep-SNN, RATE-T8 and TTFS-T8 reach 88.61\% \ensuremath{\pm} 1.77\% and 86.94\% \ensuremath{\pm} 1.25\%, respectively. Their paired mean differences from DIRECT are \ensuremath{-}0.23 pp (95\% interval {[}\ensuremath{-}2.34, 1.88{]}) and \ensuremath{-}1.90 pp ({[}\ensuremath{-}3.29, \ensuremath{-}0.50{]}) (Table~\ref{tab:3}). RATE-T8 increases the complete proxy by 9.84\%, whereas TTFS-T8 reduces it by 7.39\% with a consistent accuracy loss across the three seeds. INPUT-PG has the lowest complete proxy among the five Deep-SNN variants. These comparisons use the finite input implementations defined in the protocol.

\begin{table}[!htbp]
\centering
\caption{Software results after 50-epoch training over seeds 7, 17, and 27. Accuracy uses validation-selected checkpoints; activity counts are means per clip over the three seeds. RATE-T8 and TTFS-T8 are input-only Deep-SNN baselines. Input/coded event counts follow the definitions in Methods, and the complete proxy is the cross-variant workload measure.}
\label{tab:2}
\begingroup
\fontsize{8.7}{10.6}\selectfont
\setlength{\tabcolsep}{3.3pt}
\renewcommand{\arraystretch}{1.2}
\begin{tabular}{@{}>{\raggedright\arraybackslash}p{0.12000\dimexpr\linewidth-12\tabcolsep\relax}>{\raggedright\arraybackslash}p{0.16500\dimexpr\linewidth-12\tabcolsep\relax}>{\raggedleft\arraybackslash}p{0.16000\dimexpr\linewidth-12\tabcolsep\relax}>{\raggedleft\arraybackslash}p{0.14500\dimexpr\linewidth-12\tabcolsep\relax}>{\raggedleft\arraybackslash}p{0.13000\dimexpr\linewidth-12\tabcolsep\relax}>{\raggedleft\arraybackslash}p{0.14500\dimexpr\linewidth-12\tabcolsep\relax}>{\raggedleft\arraybackslash}p{0.13500\dimexpr\linewidth-12\tabcolsep\relax}@{}}
\toprule
\textbf{Architecture}
 & \textbf{Variant}
 & \textbf{Accuracy (\%)}
 & \textbf{Input/coded events}
 & \textbf{Hidden spikes}
 & \textbf{Complete proxy}
 & \(\Delta\) \textbf{proxy}
 \\
\midrule
Deep-SNN & DIRECT & \(88.84 \pm 1.55\) & 0.0 & 19,344.3 & 29,276.8 & 0.00\% \\
 & \textbf{INPUT-PG~(ours)} & \textbf{\boldmath \(88.47 \pm 3.70\)} & \textbf{2,806.5} & \textbf{20,595.0} & \textbf{23,401.5} & \textbf{\boldmath \(\mathbf{- 20.07\%}\)} \\
 & ALL-PG~\textbf{(ours)} & \(88.66 \pm 0.45\) & 27,788.7 & diagnostic & 27,788.7 & \(- 5.08\%\) \\
 & RATE-T8 & 88.61 \ensuremath{\pm} 1.77 & 12,867.6 & 19,290.2 & 32,157.8 & +9.84\% \\
 & TTFS-T8 & 86.94 \ensuremath{\pm} 1.25 & 7,610.9 & 19,502.7 & 27,113.6 & \ensuremath{-}7.39\% \\
LIF-TCN & DIRECT & \(88.75 \pm 2.57\) & 0.0 & 26,283.6 & 36,216.1 & 0.00\% \\
 & \textbf{INPUT-PG~(ours)} & \textbf{\boldmath \(88.10 \pm 0.21\)} & \textbf{2,684.8} & \textbf{26,381.5} & \textbf{29,066.3} & \textbf{\boldmath \(\mathbf{- 19.74\%}\)} \\
 & ALL-PG~\textbf{(ours)} & \(86.85 \pm 1.45\) & 30,289.9 & diagnostic & 30,289.9 & \(- 16.36\%\) \\
\bottomrule
\end{tabular}
\endgroup
\end{table}

\begin{table}[!htbp]
\centering
\caption{Paired accuracy differences relative to DIRECT. Intervals are 95\% paired Student-\(t\) intervals over three seeds and are reported in percentage points.}
\label{tab:3}
\begingroup
\fontsize{9.4}{11.6}\selectfont
\setlength{\tabcolsep}{3.3pt}
\renewcommand{\arraystretch}{1.2}
\begin{tabular}{@{}>{\raggedright\arraybackslash}p{0.21000\dimexpr\linewidth-4\tabcolsep\relax}>{\raggedright\arraybackslash}p{0.34000\dimexpr\linewidth-4\tabcolsep\relax}>{\raggedleft\arraybackslash}p{0.45000\dimexpr\linewidth-4\tabcolsep\relax}@{}}
\toprule
\textbf{Architecture}
 & \textbf{Comparison}
 & \textbf{Mean {[}95\% CI{]}}
 \\
\midrule
Deep-SNN & \textbf{INPUT-PG} \textbf{(ours)}\textbf{--DIRECT} & \textbf{\boldmath \ensuremath{-}0.37 {[}\ensuremath{-}5.73, 4.99{]}} \\
Deep-SNN & ALL-PG \textbf{(ours)}--DIRECT & \ensuremath{-}0.19 {[}\ensuremath{-}3.99, 3.62{]} \\
Deep-SNN & RATE-T8--DIRECT & \ensuremath{-}0.23 {[}\ensuremath{-}2.34, 1.88{]} \\
Deep-SNN & TTFS-T8--DIRECT & \ensuremath{-}1.90 {[}\ensuremath{-}3.29, \ensuremath{-}0.50{]} \\
LIF-TCN & \textbf{INPUT-PG} \textbf{(ours)}\textbf{--DIRECT} & \textbf{\boldmath \ensuremath{-}0.65 {[}\ensuremath{-}6.73, 5.43{]}} \\
LIF-TCN & ALL-PG \textbf{(ours)}--DIRECT & \ensuremath{-}1.90 {[}\ensuremath{-}8.27, 4.48{]} \\
\bottomrule
\end{tabular}
\endgroup
\end{table}

The boundary counts explain why the benefit is concentrated at the input. In the Deep-SNN, the input boundary emits 2,917.9 events per clip, whereas the three hidden boundaries emit 10,659.0, 10,735.7, and 3,476.2 events. In the LIF-TCN, the four block boundaries emit 5,432.8, 5,059.0, 6,016.2, and 10,808.8 events. These hidden boundaries already carry binary, sparse LIF activity, so encoding their transitions can add packet traffic instead of removing redundancy. This pattern accounts for the higher complete proxy of ALL-PG and supports placing PG at the acoustic input. Figure~\ref{fig:3}(c-d) shows the counts.

\subsection{Routed energy}\label{routed-energy}

For one fixed clip after Sky130HD routing, INPUT-PG reduces inference cycles by 33.76\% and SRAM reads by 34.70\% (Table~\ref{tab:4}). Its instantaneous inference dynamic power is 1.23\% higher than DIRECT, but resident and transported energy after idle subtraction decrease by 39.49\% and 39.44\%, respectively. ALL-PG also consumes less energy than DIRECT, although its reduction is smaller than that of INPUT-PG. The energy saving follows the shorter execution window and lower memory traffic, which determine the integrated energy. Figure~\ref{fig:4}(A-E) shows the estimates.

\begin{table}[!htbp]
\centering
\caption{Deep-SNN results after routing for seed 17, frozen clip 3826 (232 frames). Sky130HD TT, 25 \ensuremath{^{\circ}}C, 1.8 V, nominal 10 MHz. Energy is incremental energy after subtracting the idle baseline; Resident excludes weight load and Transported includes it. RATE-T8 and TTFS-T8 use the implementations applied only at the input, as defined in the experimental protocol.}
\label{tab:4}
\begingroup
\fontsize{8.7}{10.6}\selectfont
\setlength{\tabcolsep}{3.3pt}
\renewcommand{\arraystretch}{1.2}
\begin{tabular}{@{}>{\raggedright\arraybackslash}p{0.16000\dimexpr\linewidth-12\tabcolsep\relax}>{\raggedleft\arraybackslash}p{0.14000\dimexpr\linewidth-12\tabcolsep\relax}>{\raggedleft\arraybackslash}p{0.15500\dimexpr\linewidth-12\tabcolsep\relax}>{\raggedleft\arraybackslash}p{0.12500\dimexpr\linewidth-12\tabcolsep\relax}>{\raggedleft\arraybackslash}p{0.14500\dimexpr\linewidth-12\tabcolsep\relax}>{\raggedleft\arraybackslash}p{0.14000\dimexpr\linewidth-12\tabcolsep\relax}>{\raggedleft\arraybackslash}p{0.13500\dimexpr\linewidth-12\tabcolsep\relax}@{}}
\toprule
\textbf{Variant}
 & \textbf{Cycles}
 & \textbf{SRAM reads}
 & \textbf{Inference dyn. (mW)}
 & \textbf{Resident (\ensuremath{\mu}J)}
 & \textbf{Transported (\ensuremath{\mu}J)}
 & \textbf{Area (mm\textsuperscript{2})}
 \\
\midrule
DIRECT & 11,792,327 & 5,737,243 & 273.645 & 33,252.992 & 33,290.587 & 43.785 \\
\textbf{INPUT-PG~(ours)} & \textbf{7,810,745} & \textbf{3,746,452} & \textbf{277.010} & \textbf{20,122.617} & \textbf{20,159.901} & \textbf{44.222} \\
ALL-PG~\textbf{(ours)} & 9,158,879 & 4,420,519 & 278.367 & 24,943.967 & 24,979.233 & 44.369 \\
RATE-T8 & 16,155,227 & 7,918,693 & 269.362 & 44,365.536 & 44,400.672 & 43.638 \\
TTFS-T8 & 10,639,233 & 5,160,696 & 270.449 & 28,369.628 & 28,405.549 & 43.704 \\
\bottomrule
\end{tabular}
\endgroup
\end{table}

INPUT-PG adds about 1.00\% area and ALL-PG adds 1.33\% relative to DIRECT. Every variant transports the same 220,948 byte weight image, so the load phase changes little; the difference arises during inference. This separates the energy benefit from a weight-loading effect.

For the same fixed clip, RATE-T8 increases cycles and resident energy by 37.00\% and 33.42\% relative to DIRECT, whereas TTFS-T8 reduces them by 9.78\% and 14.69\%. INPUT-PG has the lowest resident incremental energy among the five implementations in this frozen physical experiment. Integer accuracy is reported in Tables~\ref{tab:2} and~\ref{tab:6}.

\subsection{Movable-SRAM trade-off}\label{movable-sram-trade-off}

The movable-SRAM study uses a separate \(40 \rightarrow 64 \rightarrow 5\) architecture and reports its own accuracy--energy frontier. Previous-K6 PG uses the fewest cycles and the lowest energy in this architecture, but its 79.72\% accuracy is 3.61 percentage points below DIRECT.

\begin{table}[!htbp]
\centering
\caption{Movable-SRAM experiment with a small network. Q9 accuracy uses 720 test clips; cycles and energy use one frozen clip in the movable-SRAM architecture.}
\label{tab:5}
\begingroup
\fontsize{9.4}{11.6}\selectfont
\setlength{\tabcolsep}{3.3pt}
\renewcommand{\arraystretch}{1.2}
\begin{tabular}{@{}>{\raggedright\arraybackslash}p{0.24000\dimexpr\linewidth-8\tabcolsep\relax}>{\raggedleft\arraybackslash}p{0.19000\dimexpr\linewidth-8\tabcolsep\relax}>{\raggedleft\arraybackslash}p{0.19000\dimexpr\linewidth-8\tabcolsep\relax}>{\raggedleft\arraybackslash}p{0.19000\dimexpr\linewidth-8\tabcolsep\relax}>{\raggedleft\arraybackslash}p{0.19000\dimexpr\linewidth-8\tabcolsep\relax}@{}}
\toprule
\textbf{Variant}
 & \textbf{Q9 acc. (\%)}
 & \textbf{Cycles}
 & \textbf{Resident (\ensuremath{\mu}J)}
 & \textbf{Transported (\ensuremath{\mu}J)}
 \\
\midrule
DIRECT & 83.33 & 1,374,251 & 216.633 & 216.679 \\
RATE-T8 & 81.53 & 2,443,449 & 337.234 & 337.282 \\
TTFS-T8 & 82.92 & 1,102,571 & 179.086 & 179.131 \\
\textbf{Previous-K6 PG~(ours)} & \textbf{79.72} & \textbf{703,803} & \textbf{83.651} & \textbf{83.692} \\
\bottomrule
\end{tabular}
\endgroup
\end{table}

\subsection{Integer inference accuracy and workload}\label{integer-inference-accuracy-and-workload}

With the frozen seed-17 weights, integer inference on all 720 test clips yields 90.00\% accuracy for INPUT-PG (648 clips) and 89.58\% for both DIRECT and ALL-PG (645 clips each) (Table~\ref{tab:6}). Relative to DIRECT, INPUT-PG reduces mean input activity by 71.67\%, mean SRAM reads by 33.87\%, and mean inference cycles by 32.77\%. Thus, in this software evaluation, input prediction preserves the observed task accuracy while lowering the measured workload.

\begin{table}[!htbp]
\centering
\caption{Integer software inference on 720 test clips with frozen Deep-SNN weights from seed 17. Workload columns report means per clip and exclude parameter loading. Input activity counts dense values for DIRECT and emitted input events for PG.}
\label{tab:6}
\begingroup
\fontsize{9.4}{11.6}\selectfont
\setlength{\tabcolsep}{3.3pt}
\renewcommand{\arraystretch}{1.2}
\begin{tabular}{@{}>{\raggedright\arraybackslash}p{0.30000\dimexpr\linewidth-8\tabcolsep\relax}>{\raggedleft\arraybackslash}p{0.13000\dimexpr\linewidth-8\tabcolsep\relax}>{\raggedleft\arraybackslash}p{0.18000\dimexpr\linewidth-8\tabcolsep\relax}>{\raggedleft\arraybackslash}p{0.19000\dimexpr\linewidth-8\tabcolsep\relax}>{\raggedleft\arraybackslash}p{0.20000\dimexpr\linewidth-8\tabcolsep\relax}@{}}
\toprule
\textbf{Method}
 & \textbf{Accuracy (\%)}
 & \textbf{Input activity/clip}
 & \textbf{SRAM reads/clip}
 & \textbf{Inference cycles/clip}
 \\
\midrule
DIRECT & 89.58 & 9,932.44 & 5,054,116.01 & 10,448,419.24 \\
\textbf{INPUT-PG} \textbf{(ours)} & \textbf{90.00} & \textbf{2,813.53} & \textbf{3,342,360.28} & \textbf{7,024,907.79} \\
ALL-PG \textbf{(ours)} & 89.58 & 2,932.49 & 5,016,142.84 & 10,372,472.91 \\
\bottomrule
\end{tabular}
\endgroup
\end{table}

\begin{figure}[!htb]
\centering
\includegraphics[width=147.9mm,height=.561\textheight,keepaspectratio]{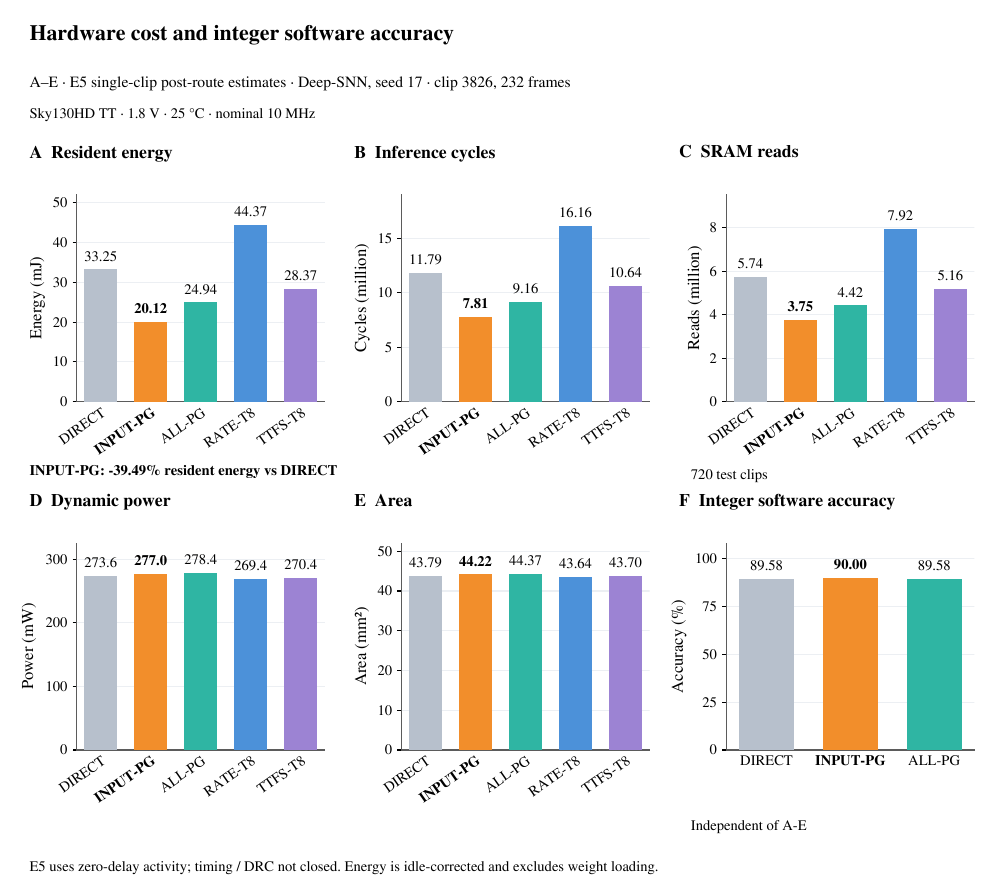}
\caption{Routed energy estimates and integer software accuracy. Panels (A-E) use Deep-SNN seed 17 and clip 3826 (232 frames) after Sky130HD routing at 1.8 V, 25 \ensuremath{^{\circ}}C and nominal 10 MHz. Resident energy is incremental dynamic energy after idle subtraction and excludes weight loading; power is active inference dynamic power. Gate activity is simulated without propagation delays, with the measured timing and DRC status reported for the prototype. Panel (F) gives integer software accuracy on 720 test clips with frozen seed-17 weights.}
\label{fig:4}
\end{figure}

The workload advantage depends on placement. ALL-PG matches the observed DIRECT accuracy but requires substantially more SRAM reads and inference cycles than INPUT-PG. The result supports applying prediction at the input instead of extending it to every hidden boundary. Figure~\ref{fig:4}(F) shows the integer accuracy results.

\section{Discussion}\label{discussion}

\subsection{Placement effects}\label{placement-effects}

The three evidence tracks point to the same placement rule. Across the two architectures, INPUT-PG reduces input activity by 71.74--72.97\% and the complete communication proxy by 19.74--20.07\%. The paired intervals are compatible with DIRECT across the three seeds, while activity reductions are consistent across architectures. With frozen integer weights, INPUT-PG reaches 90.00\% versus 89.58\% for DIRECT and reduces SRAM reads and cycles by 33.87\% and 32.77\%. In the routed run, the shorter execution window offsets the 1.23\% increase in instantaneous dynamic power. ALL-PG requires more work than INPUT-PG in both software evaluations. Together, the results support selective prediction at the input boundary.

\subsection{Workload and energy}\label{workload-and-energy}

Event counts and physical energy measure different quantities. A PG event carries an address, sign, exponent, and control information, so comparator, predictor, decoder, clock, SRAM, and routing activity all contribute to the hardware work. Instantaneous power depends on packet, predictor, decoder, clock, SRAM, and routing activity, while incremental energy also depends on execution duration. Equation~\eqref{eq:7} uses a common routed implementation, a defined phase duration, and an idle reference to calculate incremental energy. Software event counts diagnose workload reduction, whereas routed power and energy quantify the corresponding hardware cost.

\section{Conclusion}\label{conclusion}

We introduced Predictive Gating Signed-Dyadic Coding (PG), a selective input event interface that maintains a local prediction and transmits threshold-crossing signed-dyadic residuals. On the integer evaluation, INPUT-PG reached 90.00\% accuracy, reducing mean SRAM reads by 33.87\% and inference cycles by 32.77\%. For one routed clip, INPUT-PG reduced incremental inference energy by 39.49\%, excluding weight loading. Input-side predictive coding offers a hardware interface for converting temporal redundancy into lower memory and execution cost. This mechanism provides a concrete basis for research on low-power continuous speech interfaces and event-based sensory systems.

\FloatBarrier
\clearpage
\appendix
\setcounter{table}{0}
\renewcommand{\thetable}{A\arabic{table}}
\renewcommand{\theHtable}{appendix.\arabic{table}}
\FloatBarrier
\section{Detailed relative changes}\label{detailed-relative-changes}

\begin{table}[!htbp]
\centering
\caption{Deep-SNN changes relative to DIRECT for the experiment using one fixed clip in Table~\ref{tab:4}.}
\label{tab:A1}
\begingroup
\fontsize{9.4}{11.6}\selectfont
\setlength{\tabcolsep}{3.3pt}
\renewcommand{\arraystretch}{1.2}
\begin{tabular}{@{}>{\raggedright\arraybackslash}p{0.23000\dimexpr\linewidth-6\tabcolsep\relax}>{\raggedleft\arraybackslash}p{0.24000\dimexpr\linewidth-6\tabcolsep\relax}>{\raggedleft\arraybackslash}p{0.26500\dimexpr\linewidth-6\tabcolsep\relax}>{\raggedleft\arraybackslash}p{0.26500\dimexpr\linewidth-6\tabcolsep\relax}@{}}
\toprule
\textbf{Variant}
 & \textbf{Cycles}
 & \textbf{SRAM reads}
 & \textbf{Inference power}
 \\
\midrule
\textbf{INPUT-PG} \textbf{(ours)} & \textbf{\boldmath \(- 33.76\%\)} & \textbf{\boldmath \(- 34.70\%\)} & \textbf{\boldmath \(+ 1.23\%\)} \\
ALL-PG \textbf{(ours)} & \(- 22.33\%\) & \(- 22.95\%\) & \(+ 1.73\%\) \\
RATE-T8 & +37.00\% & +38.02\% & -1.57\% \\
TTFS-T8 & -9.78\% & -10.05\% & -1.17\% \\
\bottomrule
\end{tabular}
\endgroup
\end{table}

\begin{table}[!htbp]
\centering
\caption{Mean paired accuracy differences for PG placements (percentage points), summarized from Table~\ref{tab:3}.}
\label{tab:A2}
\begingroup
\fontsize{9.4}{11.6}\selectfont
\setlength{\tabcolsep}{3.3pt}
\renewcommand{\arraystretch}{1.2}
\begin{tabular}{@{}>{\raggedright\arraybackslash}p{0.25000\dimexpr\linewidth-4\tabcolsep\relax}>{\raggedleft\arraybackslash}p{0.37500\dimexpr\linewidth-4\tabcolsep\relax}>{\raggedleft\arraybackslash}p{0.37500\dimexpr\linewidth-4\tabcolsep\relax}@{}}
\toprule
\textbf{Architecture}
 & \textbf{INPUT-PG} \textbf{(ours)}\textbf{--DIRECT}
 & \textbf{ALL-PG} \textbf{(ours)}\textbf{--DIRECT}
 \\
\midrule
Deep-SNN & \textbf{\boldmath \(- 0.37\)} & \(- 0.19\) \\
LIF-TCN & \textbf{\boldmath \(- 0.65\)} & \(- 1.90\) \\
\bottomrule
\end{tabular}
\endgroup
\end{table}

\FloatBarrier

\end{document}